\documentclass[ aps preprint, 12pt,preprintnumbers]{revtex4-1}

\usepackage{graphicx}
\usepackage{epstopdf}

\usepackage{fancyhdr}
\begin{document}

\preprint{CALT-TH-2014-142}
\preprint{HDP: 14 -- 03}

\title{Banjo timbre from string stretching and frequency modulation}

\author{David Politzer}

\affiliation{California Institute of Technology}

\date{\today}

\begin{abstract} 

The geometry of a floating bridge on a drumhead soundboard produces string stretching that is first order in the amplitude of the bridge motion.  This stretching modulates the string tension and consequently modulates string frequencies at acoustic frequencies.  Early work in electronic sound synthesis identified such modulation as a source of bell-like and metallic timbre.  And increasing string stretching by adjusting banjo string-tailpiece-head geometry is known to enhance characteristic banjo tone.  Hence, this mechanism is likely a significant source of the ring, ping, clang, and plunk common to the family of instruments that share floating-bridge/drumhead construction.  Incorporating this mechanism into a full, realistic model calculation remains an open challenge.

\bigskip

\bigskip

\bigskip

**************************

\bigskip

\bigskip


\bigskip

\bigskip

**************************

\bigskip

\bigskip

\bigskip
\bigskip

\bigskip

\bigskip
\bigskip

\bigskip

\bigskip

\noindent {\bf key words:} banjo, frequency modulation, floating bridge, tailpiece

\bigskip


\bigskip

\noindent {\bf contact info:} politzer@theory.caltech.edu, (626) 395-4252, FAX: (626) 568-8473; \newline 452-48 Caltech, Pasadena CA 91125

\bigskip

\noindent {\bf separate figures:} FIG1-politzer.eps, 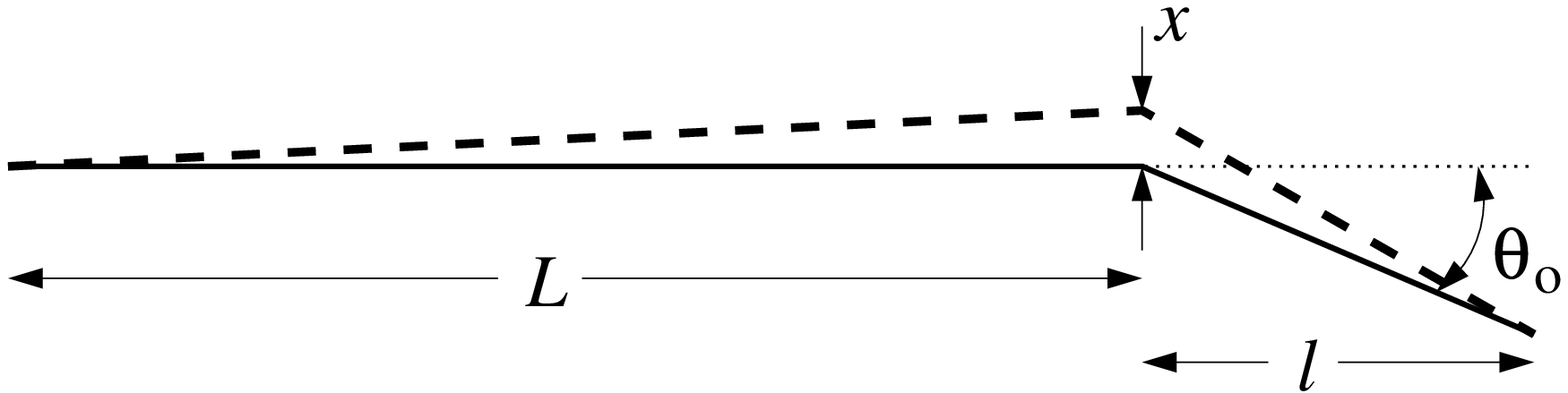, 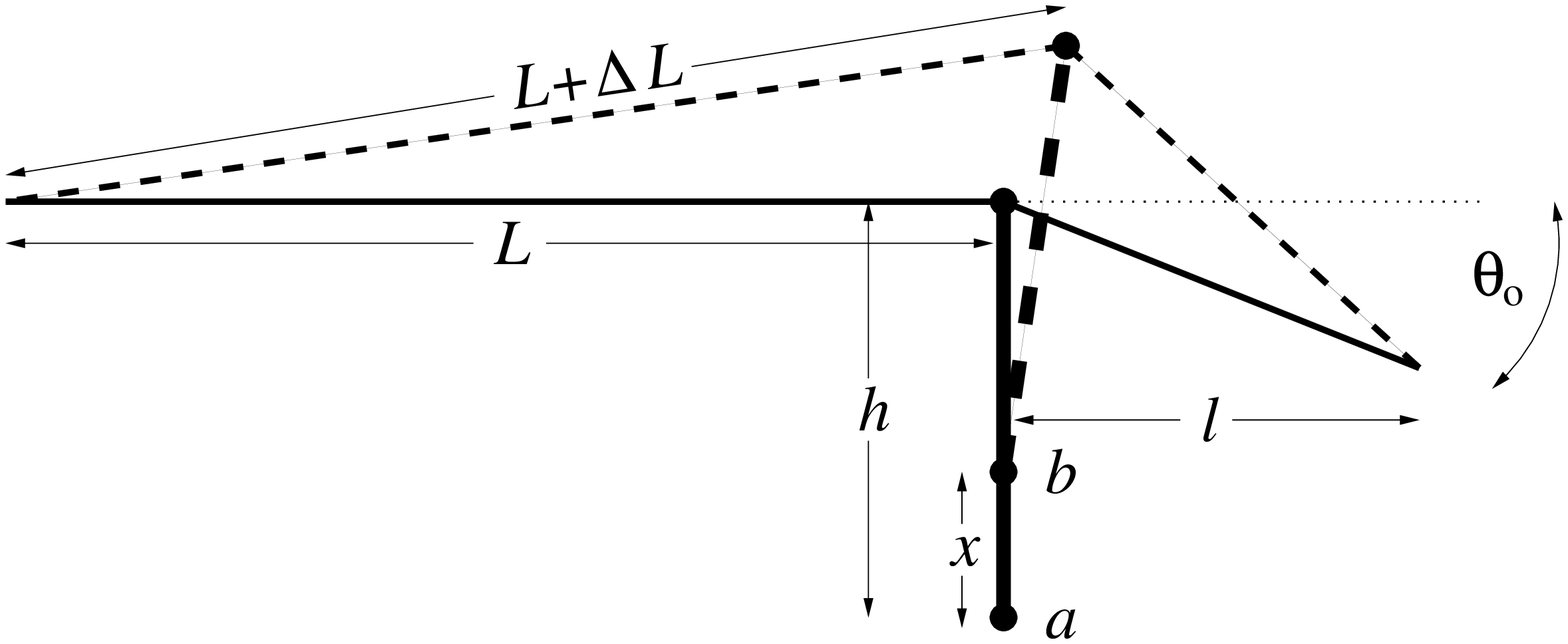

\end{abstract}

\maketitle{}

\newpage

{\bf 1. What is a Banjo?}

A banjo is a drum with strings mounted on a neck.  With minor caveats, that is what makes it a banjo.  So that is what must be responsible for its characteristic sound.  Actually, the banjo is the American instrument fitting that description.\cite{dickey},\cite{rae},\cite{moore}  Cultures around the world have their own versions.  While there is great variation among their voices, they are acoustically identifiable as belonging to the banjo family.  Among the many are the akonting and kora of west Africa, the sarod of India and its neighbors, the dramyin of Tibet, the dashpuluur of Tuva, the sanxian of China, and the shamisen of Japan.  And banjos in America today come in several readily identifiable and acoustically distinguishable varieties.

A reasonable question is: what is it in the mechanics of sound production by drum and strings that distinguishes the sound of banjos as a class from that of other stringed instruments?  While it may not be easy to quantify the defining characteristics of that sound, ``Ring the banjo" is a phrase used and commonly understood in America since before the mid-19$^{\text{th}}$ Century, an era when banjos had no metal parts.

\bigskip

{\bf 2. Geometry of Break Angle and String Stretch}

A possible answer lies in the geometry, common to all members of the banjo family,  of how the strings are attached, how they go over the bridge, and how the bridge moves.

The ideal, textbook string with fixed ends must stretch as it vibrates.  However, the amount of stretch is second order in the amplitude of vibration.  The typical textbook analysis ignores this stretching and arrives at a description of normal modes and frequencies that gives a very satisfactory account for most musical situations.  Of course, it is possible to pluck a string with such ferocity that the initial sound is, in fact, manifestly distorted by the stretching.  Even under normal conditions, second order stretching certainly contributes to the characteristic timbre of plucked strings.  Such timbre distinctions are generally very sensitive to non-linearities (e.g., as produced by stretching) and non-harmonic frequency ratios (e.g., as produced by inherent string stiffness).  However, second-order stretching and string stiffness are features common to all plucked instruments.  So they are not likely candidates for distinguishing the sound of one instrument from another, e.g., banjo from guitar.  And this would be true even if the floating bridge effects described here are in some sense smaller than the non-ideal string features common to all plucked instruments.

The floating bridge on a drumhead produces a different behavior with respect to stretch.  ``Floating" refers to the bridge's relation to the strings.  Specifically, the floating bridge goes up and down relative to the ends of the strings, which are fixed to the rim and the neck.  That is to be contrasted, for example, with a bridge and saddle, as on a flat-top guitar, where the bridge end of the string goes up and down with the bridge.

``Break angle" is the angle the strings make going over the bridge.   It is determined by the bridge height and tailpiece geometry, as roughly illustrated in FIG. 1.
(Here and in what follows, American banjo terminology is used to describe the various parts and motions.  However, all instruments in this world-wide family have analogous parts, e.g., some way to do the same job as the tailpiece to anchor the string ends to the edge of the drum.)

String tension is determined by scale length, string gauge, and chosen pitch of the open string.  With a non-zero break angle, the string tension produces a downward force on the bridge.   When the bridge is at rest, this is canceled by the upward force of the distorted head.

That there must be some string stretch somewhere is suggested by the following very simple, heuristic consideration.
In FIG. 2,  $L$ is the scale length (bridge to nut), $l$ is the bridge to tailpiece distance, and $\theta_o$ is the equilibrium break angle.  The equilibrium length of the string from nut to tailpiece is 

\centerline{$S_o = L + l$ / cos $\theta_o$  .}

\noindent If the bridge moves up a distance $x$, the total string must stretch a length

\centerline{$\Delta S =  \sqrt{L^2 + x^2} - L + \sqrt{ l^2 + (l \text{ tan} \theta_o + x)^2} - \sqrt{l^2 + (l \text{ tan} \theta_o )^2}$ .}

\noindent In practice $x$ is much smaller than $l$.  For example, $x$ could be 0.1 mm and $l$ could be 4 cm.  Using $x \ll l$:

\centerline{$\Delta S \simeq  x \text{ sin }\theta_o $ .}

\noindent As $\theta_o \to 0$ (and $x \ll l$) there remains a stretch  proportional to $x^2$, i.e., yet smaller by a factor of $x/l$.

A more realistic calculation is presented in the Appendix, which takes account of the fact that, in practice, friction prevents the strings from sliding through the bridge notches for the small motions associated with actual playing.  The stretch $\Delta L$ of the long string segment ($L$ in FIG. 2 and 3), is still first order in the vertical bridge displacement, with the bridge necessarily rocking back along the string direction in response to the vertical motion of its base.  In the limit of large stretching modulus, the stretched equilibrium condition is particularly simple.  String stretching on both sides of the bridge produces additional horizontal forces on the top of the bridge that must balance.  The balance due just to that stretching yields

\centerline{$\Delta L \simeq  x$ {\Large $\left\{ {\text{ sin }\theta_o \text{ cos }\theta_o \over 1 + \text{cos}^2 \theta_o}\right\}$}. } 

\bigskip

{\bf 3. From Stretch to Frequency Modulation}

Localized stretch and changes in tension propagate along a string at the longitudinal speed of sound in the material.  For steel strings, that is roughly 20 times greater than the speed of transverse waves in normally tuned strings.  Hence, it is reasonable to approximate the stretch as producing an instantaneous increase in tension.  If a given stretch were applied once and for all, there would be a corresponding rise in pitch.  If the stretching happened very slowly, one could still think of the stretch as a change in pitch, i.e., an adiabatic change.

Strings of different materials have different stretching moduli.  In particular, steel strings are much stiffer (longitudinally) than gut, nylon, or other synthetics.    Since it is the drumhead that moves air, the sound volume is a function of the magnitude of the bridge motion.  So, for a given sound volume, steel strings experience greater changes in tension than non-metallic strings.  In the early $20^{\text{th}}$ Century, most banjo players embraced metal strings --- for producing a sound that was more satisfyingly banjo-like  (although there have always been individuals who prefer the older and more mellow sound).  And this is a potential clue: longitudinal string stiffness is a likely contributor to banjo timbre.

Strings of different gauges mounted on a particular banjo will experience different changes in tension for a given bridge motion.  However, the fractional pitch changes will be about the same for all strings of the same material because the tuned pitches are proportional to the square root of the ratio of tension to density.

If tension changes while a string is vibrating, although the tension change is a linear response to the small length change, the string vibration is inherently non-linear.  Some care is then required when thinking in terms of Fourier components.  In particular, it is the entire bridge motion that modulates a given string's tension.  For a typical pluck, that bridge motion is not sinusoidal or even periodic.

The important picture to take from this discussion is that each string's tension is modulated by the motion of the bridge, and that motion is roughly periodic with the period of the lowest notes being played but, in fact, mirrors the full sound of the instrument.  The thus modulated tensions manifest acoustically because each string's frequencies, harmonics, and partials are proportional to the square root of its modulated tension. 

\bigskip

{\bf 4. The Sound of Frequency Modulation}

Slow frequency modulation gives a familiar form of tremolo.  In 1973, Chowning found that, when the frequency of the modulation is increased and itself enters the audio range, the tremolo warble disappears, and it is the timbre of the note that is effected.\cite{chowning}  The originally dull, sinusoidal, signal-generator  sound becomes brighter, more metallic, and bell-like when subjected to audio range frequency modulation.  The abstract mathematics is the same as for FM radio signals.\cite{FM}  In its simplest form, the modulation induces frequency sidebands along with the original signal, spaced on the order of the modulation frequency.  From ref.~\onlinecite{FM}, ``As the index [i.e., the relative frequency range of the modulation] sweeps upward, energy is swept gradually outward into higher order side bands; this is the originally exciting, now extremely annoying `FM sweep'. The important thing to get from these Bessel functions is that the higher the index, the more dispersed the spectral energy --- normally a brighter sound."

One might wonder whether string stretch from bridge motion can actually alter the sound appreciably, thinking that it cannot introduce frequencies that were not already present in its absence.  The concern is the following.  If a particular string's motion is {\it exactly} periodic and bridge motion is caused {\it only} by that string, then the frequencies of all partials are integer multiples of the fundamental frequency, with or without stretching.  Of course, on a real instrument, plucked string motion is not exactly periodic.  But, more importantly, the timbre is not just the list of frequencies present but also their relative strengths.  And the mechanism described can redistribute those strengths --- because it is non-linear.  If one imagines deconstructing the sound and then synthesizing it with an independently variable frequency modulation, nothing special happens when the modulation passes through an exact integer divisor of the frequency in question.

\bigskip

{\bf 5. Observational Support}

The proposed mechanism is inherently non-linear.  So a necessarily but not sufficient corollary is that its effects be amplitude dependent.  Indeed, banjos sound more banjo-like played loud than soft, even when the soft is put through a linear amplifier.  The clearest difference comes in the early part of the note, i.e., when the amplitude is greatest, both in the discerned sound and in the analyzed waveform and spectrum.  A careful study of the early part of each pluck of loud versus soft could confirm a non-linear origin of the banjo timbre.  However, that does not distinguish between various possible non-linear mechanisms.

Conversely, it is possible that characteristic banjo timbre results from a particularly strong linear effect that produces dramatic inharmonicity.  Strong string-drumhead coupling has been suggested independently by several people.  However, a quick comparison of banjo versus acoustic guitar using 0.010$''$ steel strings showed that the standard deviation of the first fifteen harmonic frequencies from pure integer ratios were both about 0.10\% of the average values.  The banjo was about 0.101\%, and the guitar about 0.09\%.  Perhaps more precision is needed here. 

A linear mechanism which is clearly stronger on the banjo than on other plucked string instruments is the sympathetic vibration of one string with another of their unison harmonics.  Typically, one of the strongest is the third harmonic of one string with the second of another, tuned a fifth higher in pitch.  While there does not yet exist much documented scientific literature on the American banjo, the effect is a standard element in the literature on the shamisen.\cite{shamisen}  This effect certainly contributes to a bright, quick sound for drumhead instruments, where bridge motion enhances that inter-string coupling.

These and other mechanisms deserve further study. 

Three kinds of readily available observations seem to support specifically the string stretching and frequency modulation proposal.  First, with modern software, you can construct functions of time and then listen to them.  In particular, you can listen to the sound of sinusoidal modulations of sinusoidal functions and even add an amplitude envelope typical of plucked string sound.\cite{risset}  Of course, it will not sound like a banjo.  A huge number of details are missing.  But the extra ring and brightness of tone stand out.

A second demonstration requires the facility to record and speed up the recording.  Play a low note on a banjo and push down periodically (perhaps 6 to 12 times per second) on the head near a foot of the bridge.  That will produce an audible frequency tremolo.  Speed up the recording until the modulation frequency is well above 20 Hz.  (36 Hz should do.)  The sound will have acquired a definite metallic plink, akin to banging on sheet metal.  This is most dramatic if the original note was quite low and the original break angle as small as possible.  (Adjustment of tailpiece or choice of tailpiece can accomplish the latter.)  This demonstration could also be performed with any low note frequency tremolo on any instrument --- except that it would miss the connection to bridge motion.

The third category of support (and most relevant to the specific, proposed mechanism) comes from very well-established, universally agreed upon lore among banjo players.  Without any agreement on why or how, experienced banjo players and builders know that break angle is an important issue.  Some tailpieces are adjustable over a range with the turn of a screw, while others produce a fixed break angle, whose value depends on the geometry of the tailpiece and the banjo on which it is mounted.  The range on current, popular instruments is roughly $6^{\text{o}}$ to $15^{\text{o}}$.

Often, the tailpiece advice comes with the observation that a steeper angle produces greater down-pressure of the strings on the bridge.  However, at equilibrium, that force is canceled by an upward force of the head.  Furthermore, the string-head system acting on the bridge supplies the same return force as a function of bridge displacement as with a shallower angle --- at least over the relevant range of angles and assuming the head force on the bridge is linear with displacement over the range of bridge motion.  (Further discussion of head linearity is given below.)  It is essential to remember that the strings are retuned to their original tensions after the tailpiece is adjusted.  If stretching were ignored, the fluctuating component of the forces on the bridge would be independent of break angle, and the value of the break angle would have no sonic impact.

So, even if the mechanism of tailpiece alteration is not widely understood, the consequence is: increasing the break angle makes the sound more banjo-like.  Words that are often used to describe the sound of larger angles are: ``sharper," ``snappier," or``brighter," while smaller break angles produce ``mellow," ``warm," or ``round tone."\cite{siminoff}  It is not that gut strings with gentle break angles are not banjo-like.  It is just that steel strings with sharp break angles are more so.

And the most apparent consequence of break angle on the mechanics of sound production is through the mechanism proposed in this note.
 
\bigskip

{\bf 6. Contrast with Other Stringed Instruments}

There are other acoustic, stringed instruments with floating bridges, where the bridge moves relative to the fixed ends of the string.  These include the violin family, mandolins, and arch-top guitars.  However, their bridges, riding on wooden soundboards, do not move nearly as much as the bridge on a banjo for the same sort of pluck.  For example, the violin, with a soundboard that has around 94\% the area of a typical banjo, produces a far quieter sound when plucked.  Also, the quintessential banjo features disappear if the skin on a banjo is replaced with wood.  Such instruments exist, made by instrument manufacturers, individual luthiers, and hobbyists.  They may be called banjos if they are strung and played like banjos, but their sound is quieter, and sustain is longer.  More significantly, it is widely acknowledged that they sound distinctly like dulcimers and not at all like banjos.

\bigskip

{\bf 7. Non-linearities from the Drumhead Itself}

Drumheads were likely initially chosen for soundboards because of the sound volume produced.  They are inordinately efficient transducers of a varying, localized force into sound.  (Just tap or gently rub a drumhead and listen.)  And banjo players generally opt for as low a mass bridge as is structurally sound.  The combined effect is that, in comparison to other stringed instruments, the banjo is relatively loud, with a short sustain.  This is certainly an essential aspect of its characteristic sound, amplitude envelope being an important part of distinguishing different sounds.\cite{risset}  However, the timbre corresponding to the banjo's ``ring" is something beyond that.

Banjo drumheads also have their own characteristic sound.  Some of that comes from their interaction with string tension {\it via} the bridge (as discussed above).  But there may well be other non-linearities inherent in the use of a drumhead that contribute to that sound, as well.  This deserves further study, but some basic issues are clear.

The dynamics of the head are relevant both for how the whole head vibrates in response to driving by the bridge and how the head pushes back on the bridge.  The non-trivial stress tensor of a banjo head, even at equilibrium, is apparent to the player, particularly in the vicinity of the bridge.   In addition, it is also possible that typical motions of the head in the vicinity of the bridge go beyond the range in which the relation of stress to strain can be linearized.  This is an additional possible mechanism for the brightening of sound by the generic drum/string system.  But modeling or even just picturing a non-linear stress/strain relation as it impacts bridge motion is far more challenging than the one-dimensional analog presented by the string.

\bigskip

{\bf 8. Further Work}

Motion of the bridge of a stringed instrument has long been a subject of study.\cite{raman},\cite{morse} The bridge end of the string must move to transfer energy.  However, incorporating the concomitant stretching required by the floating bridge is something that has not as yet been done.  This letter simply highlights the issue and identifies an obvious consequence. One way to proceed further would be to incorporate into a model the stretching string, the various forces on the bridge, and the dynamics of the head.  The resulting predicted sounds could be compared to real instruments, with particular attention to the sonic consequence of each part varied separately.  There are certainly other aspects that contribute to the characteristic sounds of different banjo designs.  But of particular interest here is the identification of what they all have in common but is unique to their family.  

\bigskip

{\bf 9. Summary}

Any specific instrument in the banjo family has features responsible for its characteristic timbre, and these vary considerably.  Also, banjo players are known for their penchant for adjusting and swapping parts in a quest for their own notion of ideal sound.  There is no agreed-upon ideal.  However, all these instruments have a pluck which identifies them as banjo-like and distinguishes them from anything else.  First-order string stretching and the consequent frequency modulation are proposed as a key contributor to that distinctive sound.

\bigskip

{\bf Acknowledgement}

I would like to thank Frank Rice of Caltech for constructive observations.

\bigskip

\newpage

{\bf Appendix: String-Stuck-to-Bridge Geometry}

In practice, the friction from down pressure of the strings on the bridge prevents them from sliding over the bridge as it goes up and down.  (Players often notice this sticking when tuning.)  Similarly, the base of the bridge is fixed by friction relative to the head.  So the frictional forces are forces of constraint, and there is actually a range of angles over which the bridge can be set relative to the head and strings.  What happens when the bridge is in motion is a complex, dynamical question that depends on the bridge mass and geometry and on the head elastic moduli, and it couples the motions of all the strings.  However, a more realistic estimate of the string stretching than the one presented in Section 2 can be made for a vertical bridge base displacement $x$, assuming that the situation is static.  For simplicity, also assume that the elastic modulus of the head is so much higher than that of the string that only bridge base motion perpendicular to the head is allowed.  (This is the standard picture of vibrating diaphragms and strings.)  If the base of the bridge is raised, the top of the bridge rocks back toward the tailpiece, but the total torque of the string segments about the bridge base must remain zero in equilibrium.  A possible geometry is sketched in FIG. 3, where the bridge is initially perpendicular to the long part of the string.  Note that for the string to give zero net torque about the bridge base (point $a$) in its initial position, the initial tension in the tailstring must be higher than in the long part by a factor 1/cos\thinspace$\theta_o$.


Using the parameters defined in FIG. 3, for small vertical bridge base motion $x$, the string stretch $\Delta L$ generically has a term linear in $x$ whose coefficient is a function of the break angle $\theta_o$, the lengths $L$ and $l$, the bridge height $h$, and the string stretching elastic constant $k$.  $k$ is proportional to the Young's modulus of the string and is the proportionality constant in $T_L = k \thinspace \Delta_o L$, where $T_L$ is the initial, tuned tension in the long string segment of stretched length $L$, and $\Delta_o L$ is the amount it had to be stretched to reach that tension.  The natural hierarchy of length scales is $\Delta L < x \ll \Delta L_o \ll h < l \ll L$.

To lowest order in $x$, the balance of torques at equilibrium implies 
\bigskip

\centerline{$\Delta L  =  x$  {\Large ${\text{sin} \theta_o \text{cos} \theta_o \over 1 + \text{cos}^2 \theta_o} \left\{ {1 \ - \ {\Delta L_o \over l}  \over 1 \ - \ { \text{tan} \theta_o \over 1 + \text{cos}^2 \theta_o} \ {\Delta L_o \over h}  (1 \ - \ \text{sin} \theta_o \text{cos} \theta_o \ h/l)} \right\}$}  .}

\bigskip
\noindent The leading term for $\Delta L_o \to 0$ (which is equivalent to $k \to \infty$) represents the balance of the additional horizontal forces on the top of the bridge due to the new stretching that accompanies $x$.  The terms that are down by $\Delta L_o$ (or $1/k$) arise because of the additional need to balance the torques on the bridge due to the original ($x=0$) tensions when the top point then moves in response to $x$.

To lowest order in $x$, the vertical motion of the top of the bridge is the same as the motion perpendicular to the long string segment, and both are equal to $x$.  So this aspect is no different from any stringed instrument.

\newpage


\bigskip

\bigskip

\bigskip

{\bf Figure Captions}

FIG 1:  schematic of break angle, tailpiece, string, bridge, and head

FIG 2: break angle $\theta_o$ and bridge motion $x$ determine stretch

FIG 3: bridge base motion $x$, with string stuck to bridge of height $h$

\newpage


\begin{figure}[h!]
\includegraphics[width=4.4in]{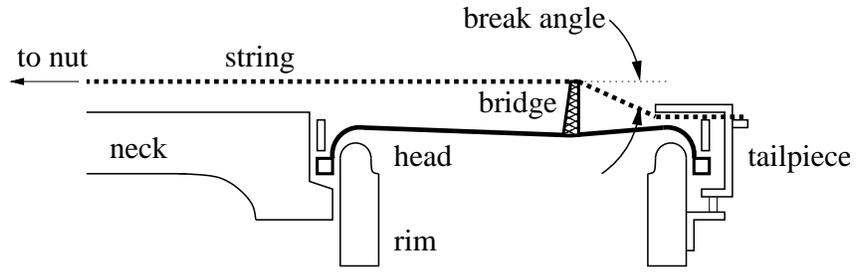}
\caption{schematic of break angle, tailpiece, string, bridge, and head }
\end{figure}

\begin{figure}[h!]
\includegraphics[width=3.7in]{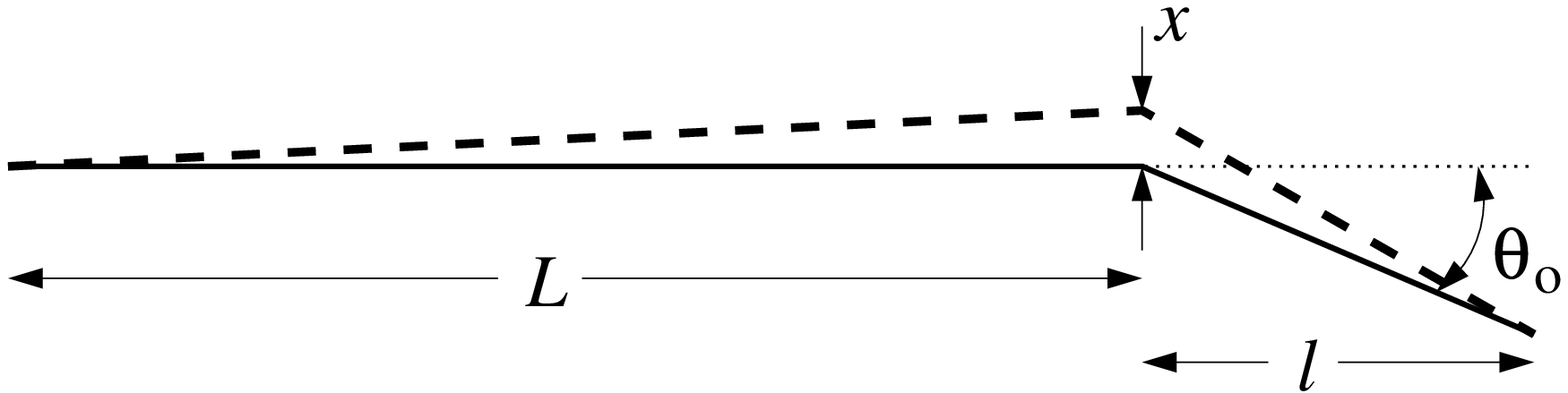}
\caption{break angle $\theta_o$ and bridge motion $x$ determine stretch}
\end{figure}

\begin{figure}[h!]
\includegraphics[width=4.3in]{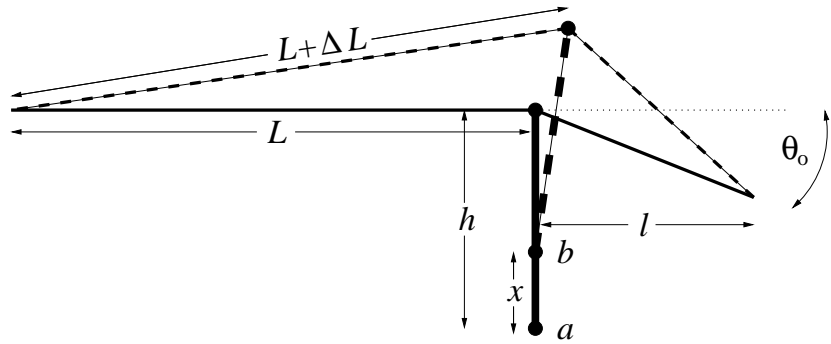}
\caption{bridge base motion $x$, with string stuck to bridge of height $h$}
\end{figure}

\end{document}